\input harvmac.tex



\lref\WittenJones{
E.~Witten, ``Quantum Field Theory And The Jones Polynomial,''
Commun.\ Math.\ Phys.\  {\bf 121}, 351 (1989).}

\lref\HOMFLY{P.~Freyd, D.~Yetter, J.~Hoste, W.~Lickorish,
K.~Millett, A.~Oceanu, ``A New Polynomial Invariant of Knots and Links,''
Bull. Amer. Math. Soc. {\bf 12} (1985) 239.}

\lref\Wittencsstring{ E.~Witten,
``Chern-Simons gauge theory as a string theory,''
Prog.\ Math.\  {\bf 133} (1995) 637, hep-th/9207094.}

\lref\OV{
H.~Ooguri, C.~Vafa, ``Knot Invariants and Topological Strings,''
Nucl.Phys. {\bf B577} (2000) 419.}

\lref\Khovanov{M.~ Khovanov,
``A categorification of the Jones polynomial,'' math.QA/9908171.}

\lref\Khovanovii{M.~ Khovanov,
``Categorifications of the colored Jones polynomial,'' math.QA/0302060.}

\lref\Khovanoviii{M.~ Khovanov,
``$sl(3)$ link homology I,'' math.QA/0304375.}

\lref\Khovanoviv{M.~ Khovanov,
``An invariant of tangle cobordisms,'' math.QA/0207264.}

\lref\DBN{D.~Bar-Natan,
``On Khovanov's categorification of the Jones polynomial,'' math.QA/0201043.}

\lref\DBNnews{D.~Bar-Natan, ``Some Khovanov-Rozansky Computations'' \semi
{http://www.math.toronto.edu/~drorbn/Misc/KhovanovRozansky/index.html}}

\lref\Jacobsson{M.~Jacobsson,
``An invariant of link cobordisms from Khovanov's homology theory,''
math.GT/0206303.}

\lref\RKhovanov{M.~Khovanov, L.~Rozansky,
``Matrix factorizations and link homology,'' math.QA/0401268.}

\lref\Rfoam{L.~Rozansky,
``Topological A-models on seamed Riemann surfaces,'' hep-th/0305205.}

\lref\KRfoam{M.~Khovanov and L.~Rozansky,
``Topological Landau-Ginzburg models on a world-sheet foam,'' hep-th/0404189.}

\lref\GopakumarV{R.~Gopakumar and C.~Vafa,
``On the gauge theory/geometry correspondence,''
Adv.\ Theor.\ Math.\ Phys.\  {\bf 3} (1999) 1415, hep-th/9811131.}

\lref\GViii{R.~Gopakumar and C.~Vafa,
``M-theory and topological strings. I,II,''
hep-th/9809187; hep-th/9812127.}

\lref\KKV{S.~Katz, A.~Klemm and C.~Vafa,
``M-theory, topological strings and spinning black holes,''
Adv.\ Theor.\ Math.\ Phys.\  {\bf 3} (1999) 1445, hep-th/9910181.}

\lref\mirbook{``Mirror Symmetry'' (Clay Mathematics Monographs, V. 1),
K.~Hori et.al. ed, American Mathematical Society, 2003.}

\lref\HV{K.~Hori, C.~Vafa, ``Mirror Symmetry,'' hep-th/0002222;
K.~Hori, A.~Iqbal, C.~Vafa, ``D-Branes And Mirror Symmetry,'' hep-th/0005247.}

\lref\AKV{M.~Aganagic, A.~Klemm, C.~Vafa,
``Disk Instantons, Mirror Symmetry and the Duality Web,'' hep-th/0105045.}

\lref\AKMV{M.~Aganagic, A.~Klemm, M.~Marino and C.~Vafa,
``Matrix model as a mirror of Chern-Simons theory,''
JHEP {\bf 0402}, 010 (2004), hep-th/0211098.}

\lref\AAHV{B.~Acharya, M.~Aganagic, K.~Hori and C.~Vafa,
``Orientifolds, mirror symmetry and superpotentials,'' hep-th/0202208.}

\lref\LMV{J.~M.~F.~Labastida, M.~Marino and C.~Vafa,
``Knots, links and branes at large N,''
JHEP {\bf 0011}, 007 (2000), hep-th/0010102.}

\lref\LMtorus{J.~M.~F.~Labastida and M.~Marino,
``Polynomial invariants for torus knots and topological strings,''
Commun.\ Math.\ Phys.\  {\bf 217} (2001) 423, hep-th/0004196.}

\lref\LMqa{J.~M.~F.~Labastida and M.~Marino,
``A new point of view in the theory of knot and link invariants,''
math.qa/0104180.}

\lref\HSTa{S.~Hosono, M.-H.~Saito, A.~Takahashi,
``Holomorphic Anomaly Equation and BPS State Counting of Rational
Elliptic Surface,'' Adv.Theor.Math.Phys. {\bf 3} (1999) 177.}

\lref\HSTb{S.~Hosono, M.-H.~Saito, A.~Takahashi,
``Relative Lefschetz Action and BPS State Counting,''
Internat. Math. Res. Notices, (2001), No. 15, 783.}

\lref\Kprivate{M.~Khovanov, private communication.}

\lref\Taubes{
C.~Taubes, ``Lagrangians for the Gopakumar-Vafa conjecture,''
math.DG/0201219.}

\lref\Wittenams{E.~Witten, ``Dynamics of Quantum Field Theory,''
{\it Quantum Fields and Strings: A Course for Mathematicians}
(P. Deligne, {\it et.al.} eds.), vol. 2, AMS Providence, RI, (1999) pp. 1313-1325.}

\lref\LVW{W.~Lerche, C.~Vafa and N.~P.~Warner,
``Chiral Rings In N=2 Superconformal Theories,''
Nucl.\ Phys.\ B {\bf 324}, 427 (1989).}

\lref\HMoore{J.~A.~Harvey and G.~W.~Moore,
``On the algebras of BPS states,''
Commun.\ Math.\ Phys.\  {\bf 197} (1998) 489, hep-th/9609017.}

\lref\IqbalV{T.~J.~Hollowood, A.~Iqbal and C.~Vafa,
``Matrix models, geometric engineering and elliptic genera,''
hep-th/0310272.}

\lref\Schwarz{A.~Schwarz,
``New topological invariants arising in the theory of quantized fields,''
Baku International Topological Conf., Abstracts (part II) (1987).}

\lref\SchwarzS{A.~Schwarz and I.~Shapiro,
``Some remarks on Gopakumar-Vafa invariants,'' hep-th/0412119.}

\lref\Aspinwallrev{ P.~S.~Aspinwall, ``D-branes on Calabi-Yau
manifolds,'' hep-th/0403166.}

\lref\MNOP{D.~Maulik, N.~Nekrasov, A.~Okounkov, R.~Pandharipande,
``Gromov-Witten theory and Donaldson-Thomas theory, I,'' math.AG/0312059.}

\lref\Katz{S.~Katz, ``Gromov-Witten, Gopakumar-Vafa, and Donaldson-Thomas
invariants of Calabi-Yau threefolds,'' math.ag/0408266.}

\lref\FultonH{W.~Fulton, J.~Harris,
``Representation Theory: A First Course,'' Springer-Verlag 1991.}

\lref\Kontsevich{M.~Kontsevich, unpublished.}

\lref\KapustinLi{A.~Kapustin and Y.~Li,
``D-branes in Landau-Ginzburg models and algebraic geometry,''
JHEP {\bf 0312} (2003) 005, hep-th/0210296.}

\lref\KapustinLii{A.~Kapustin and Y.~Li,
``Topological correlators in Landau-Ginzburg models with boundaries,''
Adv.\ Theor.\ Math.\ Phys.\  {\bf 7} (2004) 727, hep-th/0305136.}

\lref\KapustinLiii{A.~Kapustin and Y.~Li,
``D-branes in topological minimal models: The Landau-Ginzburg approach,''
JHEP {\bf 0407} (2004) 045, hep-th/0306001.}

\lref\Brunneriii{
I.~Brunner, M.~Herbst, W.~Lerche and J.~Walcher,
``Matrix factorizations and mirror symmetry: The cubic curve,''
hep-th/0408243.}

\lref\Brunner{I.~Brunner, M.~Herbst, W.~Lerche and B.~Scheuner,
``Landau-Ginzburg realization of open string TFT,'' hep-th/0305133.}

\lref\HLLii{M.~Herbst, C.~I.~Lazaroiu and W.~Lerche,
``D-brane effective action and tachyon condensation in topological minimal
models,'' hep-th/0405138.}

\lref\HLL{M.~Herbst, C.~I.~Lazaroiu and W.~Lerche,
``Superpotentials, A(infinity) relations and WDVV equations for open
topological strings,'' hep-th/0402110. }

\lref\LercheJW{W.~Lerche and J.~Walcher,
``Boundary rings and N = 2 coset models,''
Nucl.\ Phys.\ B {\bf 625} (2002) 97, hep-th/0011107.}

\lref\HoriJW{K.~Hori and J.~Walcher,
``F-term equations near Gepner points,'' hep-th/0404196.}

\lref\Orlov{D.~Orlov, ``Triangulated Categories of Singularities
and D-Branes in Landau-Ginzburg Orbifold,'' math.AG/0302304.}

\lref\Emanuelii{S.~K.~Ashok, E.~Dell'Aquila, D.~E.~Diaconescu and
B.~Florea, ``Obstructed D-branes in Landau-Ginzburg orbifolds,''
hep-th/0404167.}

\lref\Emanuel{ S.~K.~Ashok, E.~Dell'Aquila and D.~E.~Diaconescu,
``Fractional branes in Landau-Ginzburg orbifolds,''
hep-th/0401135.}

\lref\Guadagnini{E~.Guadagnini, ``The Link Invariants of the Chern-Simons
Field Theory: New Developments in Topological Quantum Field Theory,''
Walter de Gruyter Inc., 1997.}

\lref\MOY{H.~Murakami, T.~Ohtsuki, S.~Yamada,
``HOMFLY polynomial via an invariant of colored plane graphs,''
Enseign. Math. {\bf 44} (1998) 325.}

\lref\Shumakovitch{A.~Shumakovitch, {\it KhoHo} --- a program for
computing and studying Khovanov homology, {http://www.geometrie.ch/KhoHo}}

\lref\DGR{ N.~Dunfield, S.~Gukov, J.~Rasmussen, ``The
Superpolynomial for Knot Homologies,'' math.GT/0505662.}

\def\boxit#1{\vbox{\hrule\hbox{\vrule\kern8pt
\vbox{\hbox{\kern8pt}\hbox{\vbox{#1}}\hbox{\kern8pt}}
\kern8pt\vrule}\hrule}}
\def\mathboxit#1{\vbox{\hrule\hbox{\vrule\kern8pt\vbox{\kern8pt
\hbox{$\displaystyle #1$}\kern8pt}\kern8pt\vrule}\hrule}}


\let\includefigures=\iftrue
\newfam\black
\includefigures
\input epsf
\def\figin{\epsfcheck\figin}\def\figins{\epsfcheck\figins}
\def\epsfcheck{\ifx\epsfbox\UnDeFiNeD
\message{(NO epsf.tex, FIGURES WILL BE IGNORED)}
\gdef\figin##1{\vskip2in}\gdef\figins##1{\hskip.5in}
\else\message{(FIGURES WILL BE INCLUDED)}%
\gdef\figin##1{##1}\gdef\figins##1{##1}\fi}
\def\DefWarn#1{}
\def\figinsert{\goodbreak\midinsert}
\def\ifig#1#2#3{\DefWarn#1\xdef#1{fig.~\the\figno}
\writedef{#1\leftbracket fig.\noexpand~\the\figno}%
\figinsert\figin{\centerline{#3}}\medskip\centerline{\vbox{\baselineskip12pt
\advance\hsize by -1truein\noindent\footnotefont{\bf Fig.~\the\figno:} #2}}
\bigskip\endinsert\global\advance\figno by1}
\else
\def\ifig#1#2#3{\xdef#1{fig.~\the\figno}
\writedef{#1\leftbracket fig.\noexpand~\the\figno}%
\global\advance\figno by1}
\fi

\newdimen\tableauside\tableauside=1.0ex
\newdimen\tableaurule\tableaurule=0.4pt
\newdimen\tableaustep
\def\phantomhrule#1{\hbox{\vbox to0pt{\hrule height\tableaurule width#1\vss}}}
\def\phantomvrule#1{\vbox{\hbox to0pt{\vrule width\tableaurule height#1\hss}}}
\def\sqr{\vbox{%
  \phantomhrule\tableaustep
  \hbox{\phantomvrule\tableaustep\kern\tableaustep\phantomvrule\tableaustep}%
  \hbox{\vbox{\phantomhrule\tableauside}\kern-\tableaurule}}}
\def\squares#1{\hbox{\count0=#1\noindent\loop\sqr
  \advance\count0 by-1 \ifnum\count0>0\repeat}}
\def\tableau#1{\vcenter{\offinterlineskip
  \tableaustep=\tableauside\advance\tableaustep by-\tableaurule
  \kern\normallineskip\hbox
    {\kern\normallineskip\vbox
      {\gettableau#1 0 }%
     \kern\normallineskip\kern\tableaurule}%
  \kern\normallineskip\kern\tableaurule}}
\def\gettableau#1 {\ifnum#1=0\let\next=\null\else
  \squares{#1}\let\next=\gettableau\fi\next}

\tableauside=1.0ex
\tableaurule=0.4pt

\font\cmss=cmss10 \font\cmsss=cmss10 at 7pt

\def\IB{\relax\hbox{$\inbar\kern-.3em{\rm B}$}}
\def\IC{\relax\hbox{$\inbar\kern-.3em{\rm C}$}}
\def\IQ{\relax\hbox{$\inbar\kern-.3em{\rm Q}$}}
\def\ID{\relax\hbox{$\inbar\kern-.3em{\rm D}$}}
\def\IE{\relax\hbox{$\inbar\kern-.3em{\rm E}$}}
\def\IF{\relax\hbox{$\inbar\kern-.3em{\rm F}$}}
\def\IG{\relax\hbox{$\inbar\kern-.3em{\rm G}$}}
\def\IGa{\relax\hbox{${\rm I}\kern-.18em\Gamma$}}
\def\IH{\relax{\rm I\kern-.18em H}}
\def\IK{\relax{\rm I\kern-.18em K}}
\def\IL{\relax{\rm I\kern-.18em L}}
\def\IP{\relax{\rm I\kern-.18em P}}
\def\IR{\relax{\rm I\kern-.18em R}}
\def\Z{\relax\ifmmode\mathchoice
{\hbox{\cmss Z\kern-.4em Z}}{\hbox{\cmss Z\kern-.4em Z}}
{\lower.9pt\hbox{\cmsss Z\kern-.4em Z}}
{\lower1.2pt\hbox{\cmsss Z\kern-.4em Z}}\else{\cmss Z\kern-.4em
Z}\fi}

\def\II{\relax{\rm I\kern-.18em I}}

\def\S{{\bf S}}


\def\CC {{\cal C}}

\def\CH {{\cal H}}

\def\CL {{\cal L}}
\def\CM {{\cal M}}
\def\CN {{\cal N}}
\def\CO {{\cal O}}

\def\CQ {{\cal Q}}


\def\tilde{\widetilde}
\def\hat{\widehat}


\def\Tr{{\rm Tr}}
\def\Id{{\rm Id}}

\def\inbar{\,\vrule height1.5ex width.4pt depth0pt}

\def\a{\alpha}

\def\e{\epsilon}

\def\la{\lambda}

\def\lk{{\rm lk}}
\def\Tr{{\rm Tr}}

\def\IH{{\bf H}}

\def\example#1{\bgroup\narrower\footnotefont\baselineskip\footskip\bigbreak
\hrule\medskip\nobreak\noindent {\bf Example}. {\it #1\/}\par\nobreak}
\def\endexample{\medskip\nobreak\hrule\bigbreak\egroup}


\Title{\vbox{\baselineskip12pt\hbox{hep-th/0412243}
\hbox{HUTP-04/A034}
 }} {\vbox{ 
\centerline{Khovanov-Rozansky Homology and Topological Strings}
\medskip
}} 
\centerline{Sergei Gukov$^1$,
Albert Schwarz$^2$,
and Cumrun Vafa$^1$}
\medskip
\vskip 8pt
\centerline{\it $^1$ Jefferson Physical Laboratory,}
\centerline{\it Harvard University,}
\centerline{\it Cambridge, MA 02138, USA}
\medskip
\medskip
\bigskip
\centerline{\it $^2$ Department of Mathematics,}
\centerline{\it University of California,}
\centerline{\it Davis, CA 95616, USA}
\medskip

\vskip 30pt {\bf \centerline{Abstract}} \noindent

We conjecture a relation between the $sl(N)$ knot homology,
recently introduced by Khovanov and Rozansky,
and the spectrum of BPS states captured by
open topological strings.
This conjecture leads to new regularities among the $sl(N)$ knot
homology groups and suggests that they can be interpreted directly
in topological string theory.
We use this approach in various examples to predict
the $sl(N)$ knot homology groups for all values of $N$.
We verify that our predictions pass some non-trivial checks.

\smallskip
\bigskip\bigskip
\hfill{\hbox{{\it Dedicated to the memory of F.A.~Berezin.}}}
\Date{December 2004}


\newsec{Introduction and Summary}

During the past twenty years, topological field theories
have been the source of the vigorous interaction between
theoretical physics and pure mathematics,
increasingly fruitful for both fields.
One of the famous examples of topological field theories is a
Chern-Simons gauge theory \refs{\Schwarz,\WittenJones}.
 Observables in this theory are
naturally associated with knots and links.
Specifically, given an oriented  knot, $K$, and a representation of the gauge
group, $R$, one can construct a Wilson loop operator, $W_R (K)$,
whose expectation value turns out to be a polynomial invariant of $K$,
such as the Jones polynomial and its generalizations
\refs{\WittenJones}.

Here, we will be mainly interested in the case where $R=\tableau{1}$ is
the fundamental representation of $sl(N)$.
The corresponding quantum invariant
\eqn\pnjones{P_N (q) = \langle W_{\tableau{1}} (K) \rangle }
is a one-variable specialization of the HOMFLY polynomial \HOMFLY.
It can be determined by the $sl(N)$ skein relation
\eqn\skeinpn{ q^N P_N (L_+) - q^{-N} P_N (L_-) = (q^{-1} - q) P_N (L_0)}
%
and by the normalization
\eqn\njonesunknot{P_N ({\rm unknot},q) = [N] = {q^{N} - q^{-N}
\over q - q^{-1} }  }
\ifig\dervgraph{Link diagrams connected by the skein relation.}
{\epsfxsize2.7in\epsfbox{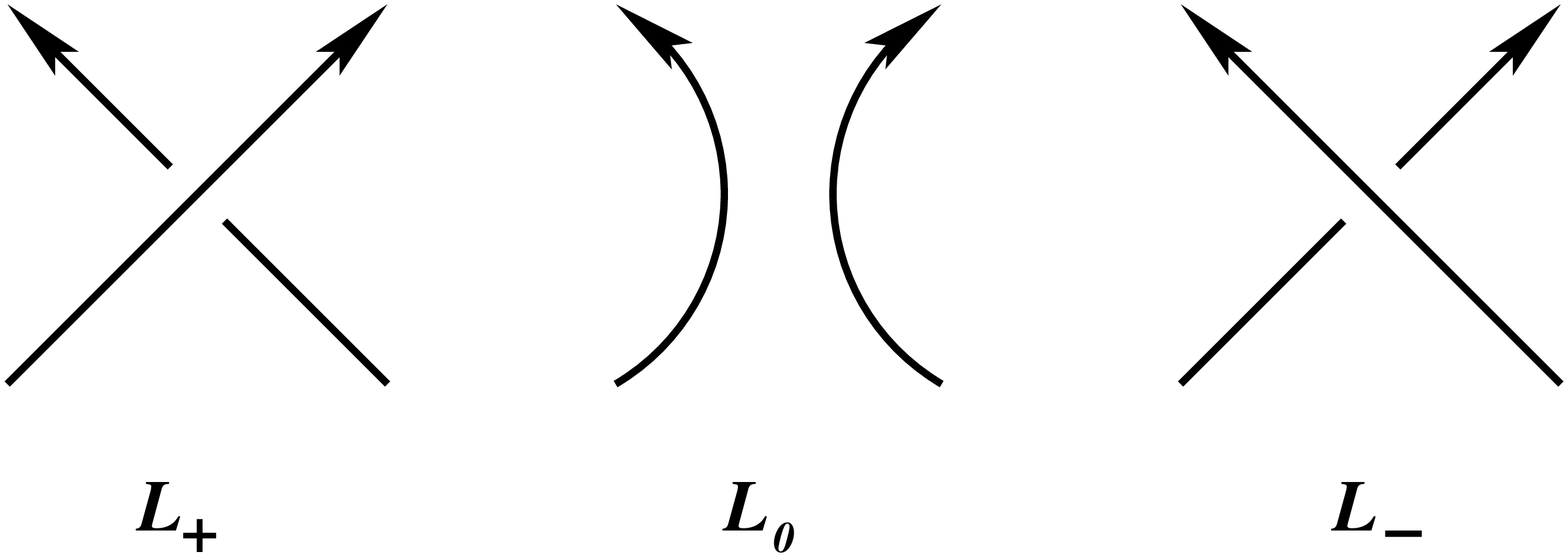}}

As shown in \Wittencsstring,
Chern-Simons theory can be embedded in string theory.
Later it was shown \refs{\GopakumarV,\GViii,\OV}\
using highly non-trivial `stringy dualities' that this leads to
a reformulation of quantum $sl(N)$ knot invariants in terms of
topological string amplitudes (Gromov-Witten invariants).
Moreover, it was realized that the polynomial invariants $P_N (q)$
can be reformulated in terms of integers $N_{\tableau{1},Q,s}$
which capture the spectrum of BPS states in the string Hilbert
space\foot{In our normalization,
the integer invariants $N_{\tableau{1},Q,s}$ are non-trivial
only for even values of $s$.} \refs{\OV,\LMV}:
\eqn\pnvian{P_N (q)= {1 \over q-q^{-1}} \sum_{s,Q \in \Z}
N_{\tableau{1},Q,s} q^{NQ+s} }
It is difficult to give a mathematically rigorous definition of
BPS degeneracies. In the simpler case of closed topological strings,
there is a similar notion of BPS degeneracies \GViii.
Attempts to give it a precise mathematical
definition \refs{\HSTa,\HSTb,\KKV,\MNOP,\Katz}
led to a deeper understanding of Gromov-Witten invariants,
though even there not all questions are answered.
In the open string case, even though a mathematically rigorous 
proof of \pnvian\ is not available
this equation is strongly supported by physical arguments and highly non-trivial checks,
in particular, by the fact that considering it as a definition of
$N_{\tableau{1},Q,s}$ one always obtains integers.

In another line of development, the quantum $sl(N)$ invariant
$P_N (q)$ was lifted to a homological knot invariant
\refs{\Khovanov,\Khovanovii,\Khovanoviii,\RKhovanov}.
For a given knot $K$ and a fixed value of $N$,
the $sl(N)$ homological invariant is a doubly graded
cohomology theory, $\CH^{i,j}_N (K)$, whose Euler characteristic with
respect to one of the gradings equals the quantum $sl(N)$ invariant,
\eqn\pnhomology{ P_N (q) = \sum_{i,j \in \Z} (-1)^i q^j
\dim \CH^{i,j}_N (K) }
Following \DBN, we also introduce the graded Poincar\'e polynomial,
\eqn\nkhpolyn{ Kh_N (q,t) := \sum_{i,j \in \Z} t^i q^j
\dim \CH_N^{i,j} (K) }
Relegating technical details to the following sections, let us summarize
some important features of $\CH_N$:

\item{$i)$}
$\CH_N$ is a functor (from the category of links and cobordisms to the category
of vector spaces \refs{\Khovanoviii,\DBN,\Khovanoviv,\Jacobsson});

\item{$ii)$}
$\CH_N$ is stronger than $P_N$, at least in the case $N=2$;

\item{$iii)$}
$\CH_N$ is hard to compute (at present, only the $sl(2)$ homological
invariant has been computed for knots with small number of
crossings \refs{\DBN,\Shumakovitch});

\item{$iv)$}
$\CH_N$ cries out for a physical interpretation!

In this paper, we take a modest step towards understanding
the physical interpretation of the $sl(N)$ knot homology by relating it
to the BPS spectrum of states in open topological strings.
It naturally leads us to new regularities among the homology
groups, $\CH^{i,j}_N (K)$, which we hope may ultimately lead
to a better understanding of the $sl(N)$ homological invariant
on the deep conceptual level.
In particular, we will be interested in the dependence
of the homology groups $\CH^{i,j}_N (K)$ on the rank of $sl(N)$.
By analogy with \pnvian,
we expect that, at least for sufficiently
large values of $N$, the following complex has a very simple structure:
%
\eqn\khshifted{ \CH^{\bullet}_N \{+1\} - \CH^{\bullet}_N \{-1\} }
where $\{n\}$ denotes the shift in $q$-grading up by $n$.
Specifically, we propose the following:

\medskip\noindent
{\bf Conjecture:} {\it For a knot $K$ and sufficiently
large\foot{It turns out, however, that for many simple
knots this conjecture works for all $N \ge 2$.}
values of $N$, the graded Poincar\'e polynomial of
the doubly graded complex \khshifted\ can be expressed
in terms of $N$-independent integer invariants $D_{Q,s,r} (K)$,}
\eqn\khvians{\mathboxit{
(q-q^{-1}) Kh_N (q,t) = \sum_{Q,s,r \in \Z} D_{Q,s,r} q^{NQ + s} t^r
}}
%
Notice, this conjecture implies that (for sufficiently large $N$)
the set of values of the grading $i$ in $\CH_N^{i,j} (K)$
is contained in a set that does not depend on $N$.
There is also a generalization of this conjecture for links, see below.

The integer invariants $D_{Q,s,r}$ provide
a refinement of the open string BPS degeneracies by decomposition
of the states in terms of
an additional $U(1)$ charge (grading) $r$ in a sense that
\eqn\nqsviad{ N_{\tableau{1},Q,s} = \sum_{r \in \Z} (-1)^r D_{Q,s,r} . }
This is parallel to the refinement of BPS degeneracies for closed
strings found in \IqbalV . 
Notice, that the conjectured relation \khvians\ is manifestly
consistent with eqs. \pnvian\ - \nkhpolyn\ and the relation
\nqsviad.

\bigskip {\noindent {\it {Organization of the Paper}}}

In section 2, we briefly review some of the main statements about
integer BPS degeneracies
for open strings and the $sl(N)$ knot homology.
The more detailed physical setup and the interpretation of the $sl(N)$ knot homology
as BPS degeneracies $D_{Q,s,r}$ is discussed in section 3.
In section 4, we study explicit examples.
In particular, we use \khvians\ to predict the $sl(N)$ homological
invariants for certain knots with small number of crossings
and verify that our predictions pass some non-trivial checks.
In section 5, we discuss various generalizations of
the above conjecture.
Finally, in the appendices we present the explicit form
of the HOMFLY polynomial and the $sl(2)$ homological
invariant for some knots.


\newsec{Preliminaries}

\subsec{Conventions}

Here we follow the standard conventions in knot theory,
where the unreduced Jones polynomial has expansion in
integer powers of $q$, $J \in \Z [q,q^{-1}]$.
For example, in these conventions, the quantum $sl(N)$ invariant
of the trivial knot has the form \njonesunknot.
These conventions are compatible with the ones used
in \refs{\Khovanov, \Khovanovii,\Khovanoviii,\RKhovanov, \DBN},
but differ from the conventions used in the physics
literature \refs{\GopakumarV,\GViii,\OV,\LMV,\LMtorus,\LMqa}.
The two sets of conventions are related by a simple change $q \to q^{1/2}$.
We also warn the reader that in some literature
our variable $q$ is denoted $q^{-1}$ and, similarly,
$t$ is denoted $t^{-1}$.

Throughout, we work over $\IQ$.
In particular, all cohomology groups are $\IQ$-vector spaces,
and $\dim (\CH^{i,j}_N)$ denotes $\dim_{\IQ} (\CH^{i,j}_N)$.


\subsec{A Brief Review of the BPS Degenaracies and Knot Invariants}

The generating function of Wilson
loop (knot) observables in Chern-Simons theory
 can be written in terms of the $f$-polynomials \OV :
\eqn\fcsviaf{
F_{CS} (V) = \sum_{d=1}^{\infty} \sum_R {1 \over d} f_R (q^d,\la^n) \Tr_R V^d
}
where $\la := q^N$ and $N$ is the rank of the gauge group.
The polynomials $f_R (q,\la)$ can be expanded in $q$ and $\la$ \refs{\OV ,\LMV}
\eqn\fRviaN{\eqalign{
f_R (q, \la) & = {1 \over q-q^{-1}} \sum_{s,Q} N_{R,Q,s} \la^Q q^s = \cr
& = \sum_{g \ge 0} \sum_Q \sum_{R' , R''}
C_{R R' R''} S_{R'} (q) \hat N_{R'', g, Q} (q^{-1} - q)^{2g-1} \la^Q
}}
In this expression, $R$, $R'$, and $R''$ denote representations
of the symmetric group, which we can label by Young tableaus
with $\ell$ boxes. The coefficients $C_{R R' R''}$ are
the Clebsch-Gordan coefficients of the symmetric group,
and the function $S_R (q)$ is non-zero only for hook
representations of the form
\eqn\hookreps{ \tableau{5 1 1 1} }
Specifically, if $R$ is a hook representation with $\ell - d$ boxes
in the first row, then $S_R (q) = (-1)^d q^{- {\ell - 1 \over 2}+d}$,
and $S_R (q) = 0$ otherwise. Finally, the numbers $N_{R,Q,s}$
and $\hat N_{R,g,Q}$ encode the BPS spectrum for open topological
strings.

The integer BPS degeneracies can be defined in terms of cohomology
computations on certain spaces.  In particular, for the fundamental represenation,
$\hat N_{\tableau{1},g,Q}$ can be defined as the Euler
characteristics of the cohomology of moduli spaces\foot{As discussed
in section 3 this is the moduli space of holomorphic curves
of genus $g$ with one boundary ending on a suitable Lagrangian submanifold.}
 $\CM_{g,Q}$. We refer the
reader to the original papers \refs{\GViii,\OV,\LMV} for the
details, and briefly sketch here the results that will be relevant
to us in what follows. There is an explicit expression for $\hat
N_{\tableau{1},g,Q}$ in terms of the cohomology of $\CM_{g,Q}$,
\eqn\gvmodulii{ \hat N_{\tableau{1},g,Q} = \e \ \chi (\CM_{g,Q})
}
where\foot{As discussed in \OV\  the origin of
the sign $\e $ is related to the analytic continuation one has
to make in order to relate topological string amplitudes with the
quantum-group knot invariants.} $\e =\pm 1$.

In the case we are considering, $R=\tableau{1}$,
the polynomial $f_{\tableau{1}}(q,\la)$ is the unnormalized
HOMFLY polynomial. Its value at $\la=q^N$
gives the $sl(N)$ polynomial $P_N (q)$:
\eqn\fwjones{ P_N (q) = f_{\tableau{1}}(q,q^N) }
The polynomial $f_{\tableau{1}}(q,\la)$
has a simple representation in terms of $\hat N_{R,g,Q}$:
\eqn\fnhat{ f_{\tableau{1}}(q,\la) = \sum_Q \sum_{g \ge 0}
\hat N_{\tableau{1},g,Q} (q^{-1} - q)^{2g-1} \la^Q
}
Comparing this formula with eq. \fRviaN, we obtain an expression
for the integers $N_{\tableau{1},Q,s}$ in terms of $\hat
N_{\tableau{1},g,Q}$,
\eqn\nvianhat{ N_{\tableau{1},Q,s} = - \sum_{g \ge 0} (-1)^{g+s/2}
{2g \choose g+s/2} \hat N_{\tableau{1},g,Q} }
Using \gvmodulii, we can further express $N_{\tableau{1},Q,s}$ via
dimensions of the cohomology groups $H^k (\CM_{g,Q})$,
\eqn\nviadimm{ N_{\tableau{1},Q,s} = - \e \sum_{g \ge 0 ,k}
(-1)^{k+g+s/2} {2g \choose g+s/2} \dim H^k (\CM_{g,Q})
 }

\example{The Unknot}
For the unknot
the corresponding moduli spaces are isolated points,
so that
\eqn\hgvunknot{ H^k (\CM_{g,Q}) = \cases{\IQ & if $k=g=0$ and
$Q=\pm 1$ \cr 0 & otherwise} }
Hence, from \gvmodulii\ we find that the only non-zero invariants
are \OV:
\eqn\nhatunknot{ \hat N_{\tableau{1},0,Q=\pm 1} = \mp 1}
This leads to the usual expression for
the HOMFLY polynomial
\eqn\funknot{
f_{\tableau{1}}({\rm unknot}) = {\la - \la^{-1} \over q - q^{-1}}
= {q^{N} - q^{-N} \over q - q^{-1}}
}
\endexample

Let us consider another example.

\example{The Trefoil Knot}
In this case, the non-zero BPS invariants are \LMV:
\eqn\nhattrefoil{\eqalign{
& \hat N_{\tableau{1},0,1} = 2, \quad
\hat N_{\tableau{1},0,3} = -3, \quad
\hat N_{\tableau{1},0,5} = 1, \quad \cr
& \hat N_{\tableau{1},1,1} = 1, \quad
\hat N_{\tableau{1},1,3} = -1
}}
Substituting this into \fnhat, we find
\eqn\ftrefoil{ f_{\tableau{1}}(3_1)
= {2 \la - 3 \la^3 + \la^5 \over q^{-1} - q} + (\la - \la^3) (q^{-1} - q)
}
Notice, that in the rank 2 case, $\la=q^2$, we recover the usual
expression for the unreduced Jones polynomial of the trefoil knot,
\eqn\jonestrefoil{ J(3_1) = q + q^3 + q^5 - q^9}
\endexample

\subsec{A Brief Review of the $sl(N)$ Knot Homology}

In a fascinating work \Khovanov, Khovanov introduced a new homological
knot invariant, which has the Jones polynomial as its graded Euler
characteristic.
Subsequently, this work was extended to a categorification of
the quantum $sl(3)$ invariant \Khovanoviii\ and, more recently,
to a categorification of the quantum $sl(N)$ invariant \RKhovanov.
Although these constructions differ in details,
the basic idea is to associate
a chain complex of graded vector spaces, $\CC^{\bullet}_N (L)$,
to a plane diagram of a link $L$ colored by a fundamental
representation of $sl(N)$.
The bigraded cohomology groups of this complex,
\eqn\hkhov{
\CH_N (L) := H^{*} (\CC^{\bullet}_N)
}
do not depend, up to isomorphism, on the choice of the projection of $L$.
It is also convenient to define the graded Poincar\'e polynomial,
\eqn\khpolyn{ Kh_N (L) := \sum_{i,j \in \Z} t^i q^j \dim \CH_N^{i,j} (L) }
and the graded Euler characteristic,
\eqn\chikhov{
\chi_q (L) := \sum_{i,j \in \Z} (-1)^i q^j \dim \CH_N^{i,j} (L) }
One of the main results in \refs{\Khovanov,\Khovanoviii,\RKhovanov}
states that $\chi_q (L)$ is equal to the quantum $sl(N)$ invariant of $L$:
\eqn\khjones{P_N (q) = \chi_q (L) = Kh_N (L) \vert_{t=-1}
}
Notice, that since the Euler characteristic of
the cohomology $\CH^{i,j}_N (L)$ is the same as
the Euler characteristic of the chain complex $\CC^{i,j}_N (L)$ itself,
we can write \chikhov\ - \khjones\ as
\eqn\pnchicc{
P_N (q) = \sum_{i,j \in \Z} (-1)^i q^j \dim \CC_N^{i,j} (L) }

\example{The Trefoil Knot} For the trefoil knot and $N=2$, the
Khovanov's invariant $Kh_{N=2} (L)$ has the following form:
\eqn\khtrefoil{Kh_{N=2} (3_1) = q + q^3 + q^5 t^2 + q^9 t^3 }
It is easy to see that at $t=-1$ we recover the usual Jones polynomial
\jonestrefoil\ of the trefoil knot.
\endexample

Now, let us say a few words about the structure of the complex
$\CC^{\bullet}_N (L)$. Let $W = \oplus_m W_m$ be a graded vector space.
Its graded dimension is
\eqn\dimqw{ \dim_q W = \sum_m q^m \dim W_m}
In the spirit of the derived categories, we will be considering
chain complexes of these graded vector spaces, {\it e.g.}
\eqn\ccomplex{
\CC^{\bullet} ~:~~~ \ldots
\longrightarrow^{\kern -20pt d^{r-1}}~~ \CC^r~~
\longrightarrow^{\kern -13pt d^{r}} ~~ \CC^{r+1}~~
\longrightarrow^{\kern -20pt d^{r+1}}~~ \CC^{r+2}~~
\longrightarrow^{\kern -20pt d^{r+2}}~~ \ldots
}
where $r$ is the ``height'' of the graded vector space $\CC^r$.

Let us also introduce two translation functors: the degree shift
\eqn\mshift{ W \{ l \}_m  := W_{m-l} }
and the height shift
\eqn\rshift{ \CC [s]^r := \CC^{r-s} }
In other words, $\CC^{\bullet} [s]$ denotes
the complex $\CC^{\bullet}$ shifted $s$ places to the right.
Also, note that \mshift\ implies
\eqn\shiftdimqw{ \dim_q W \{ l \} = q^l \dim_q W }
%


\ifig\knothoma{
A planar trivalent graph $\Gamma$ near a ``wide edge'' and
its representation in the Murakami-Ohtsuki-Yamada terminology \MOY,
where each oriented edge is labelled by a fundamental weight of $sl(N)$.}
{\epsfxsize2.5in\epsfbox{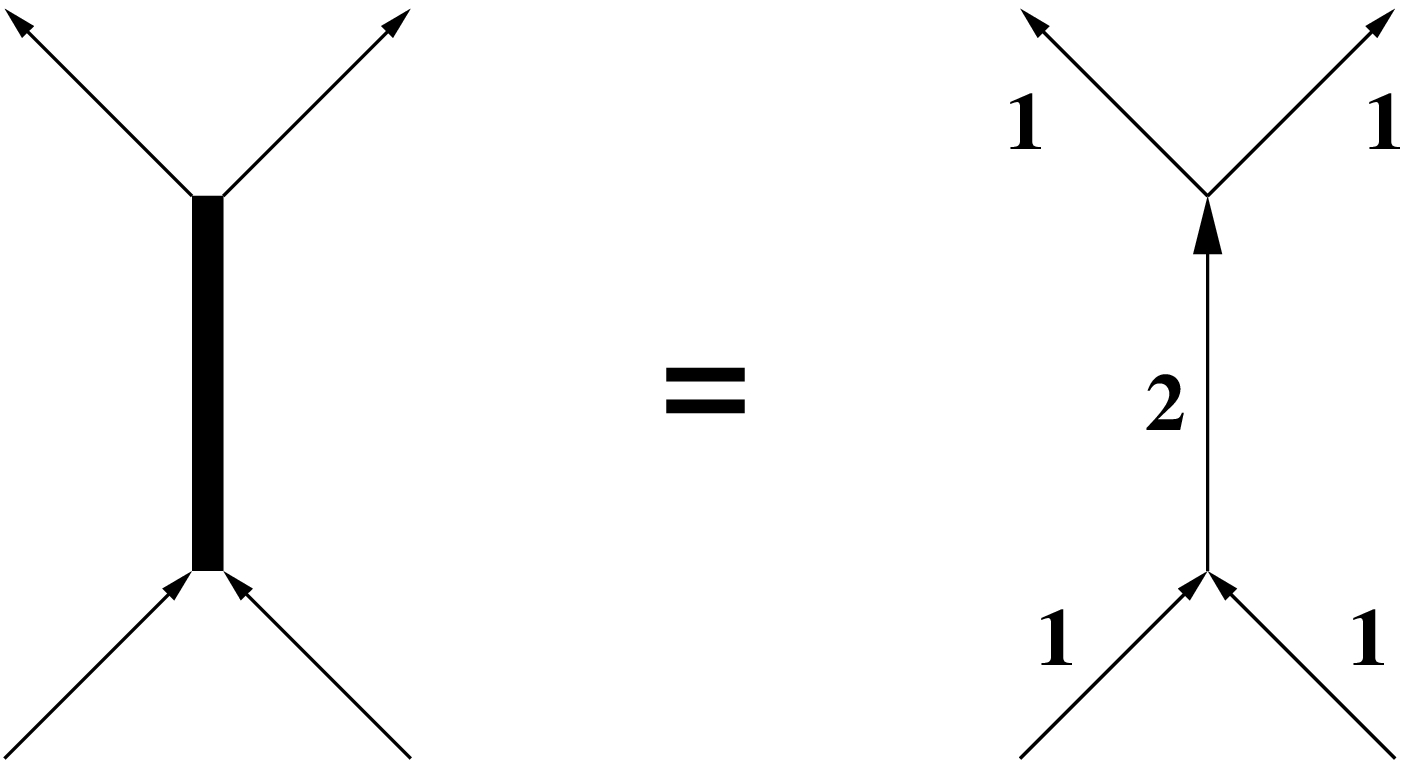}}

For each value of $N$, there is a combinatorial algorithm to
construct the $\Z \oplus \Z$-graded chain complex
$\CC_N^{\bullet} (L)$ from a plane diagram $D$ of the link $L$.
Namely, using a suitable skein relation one takes all
possible resolutions of the link diagram $D$,
\eqn\ccresol{
\CC_N^{\bullet} (L) = \bigoplus_{\Gamma} \CC_N^{\bullet} (\Gamma) }
where each resolution $\Gamma$ is a planar trivalent graph,
as in \knothoma. For example, in the $sl(2)$ case, each $\Gamma$
is a collection of plane cycles, and a simple algorithm for
constructing $\CC_2^{\bullet} (L)$ was proposed by Khovanov
in the original paper \Khovanov; essentially, one has
\eqn\sltwocc{
\CC_2^{\bullet} (\Gamma) \cong V^{\otimes \# {\rm (cycles)}} }
Here, $V$ is a cohomology ring of a 2-sphere, $V \cong H^* (\S^2)$,
so that $\dim_q (V) = q + q^{-1}$.
Similarly, in the $sl(3)$ case,
one can systematically construct $\CC_3^{\bullet} (L)$
using web cobordisms (``foams'') \Khovanoviii.

A general algorithm, which in principle\foot{
In practice, the previous two constructions remain more
suitable for explicit calculations of the $sl(2)$
and $sl(3)$ knot homology.} allows
to construct the $sl(N)$ knot homology for arbitrary $N$,
was proposed recently by Khovanov and Rozansky \RKhovanov.
In this approach, $\CC_N^{\bullet} (L)$ is constructed
as a $\Z \oplus \Z \oplus \Z_2$-graded complex
of $\IQ$-vector spaces.
It turns out that the cohomology groups of this complex are
non-trivial only for one value of the $\Z_2$ grading, so that
$H^* (\CC_N^{\bullet})$ is $\Z \oplus \Z$-graded as in \hkhov.
Basically, a 2-periodic complex $\CC_N^{\bullet} (L)$
is a tensor product of matrix factorizations:
\eqn\mdmdm{
M^0 \longrightarrow^{\kern -10pt d^0}~ M^1
\longrightarrow^{\kern -10pt d^1}~ M^0 }

We remind that a matrix factorization of
a homogeneous potential $W (x_i)$ is a collection
of two free modules $M^0$, $M^1$ over the ring
of polynomials in $x_i$, and two maps
$d^0 :~ M^0 \to M^1$ and $d^1 :~ M^1 \to M^0$,
such that
\eqn\ddfactor{d^0 d^1 = W \cdot \Id \quad , \quad d^1 d^0 = W \cdot \Id }
Choosing a basis in $M^0$, $M^1$ we can think of
$d^0$ and $d^1$ as matrices with polynomial entries.
One can also combine $d^0$ and $d^1$ into an odd matrix
of the form
\eqn\qviadd{ \CQ = \pmatrix{0 & d^1 \cr d^0 & 0 } }
so that \ddfactor\ can be written in a compact form
\eqn\wqfactor{ \CQ^2 = W \cdot \Id}
In other words, $\CQ$ can be regarded as an odd endomorphism (also
known as ``twisted differential'') acting on a free $\Z_2$-graded
module $M = M^0 \oplus M^1$ over $R = \IQ [x_i]$. When, following
Khovanov and Rozansky \RKhovanov, one takes a tensor product over
matrix factorizations of the form \mdmdm, the $W$'s cancel, so
that $\CQ$ becomes an ordinary differential, $\CQ^2=0$, and the
result becomes a complex,
\eqn\casmprod{ \CC_N^{\bullet} (L) = \otimes_i~ M_i }

A factorization $M$ is called trivial (or contractible)
if it has the form
\eqn\trivfact{ R \longrightarrow^{\kern -10pt 1}~ R
\longrightarrow^{\kern -12pt W}~ R  }
Dividing the category of matrix factorizations \mdmdm\
by the equivalence relation
\eqn\mmequiv{ M \sim M \oplus M_{{\rm trivial}} }
we get a ``derived'' category of factorizations of $W$, denoted by
${\rm hmf}_W$ in \RKhovanov, whose isomorphisms are morphisms
which induce isomorphisms in cohomology. Furthermore, this
category has a triangulated structure, which plays an important
role in the notion of D-brane stability, see \Aspinwallrev\ for a
recent review. In fact, the category ${\rm hmf}_W$ is a simple
example of a ``D-brane category'', whose objects are D-branes
in $\CN=2$ Landau-Ginzburg model with a superpotential
$W$ \refs{\Kontsevich,\KapustinLi,\Orlov,\Brunner}. In particular,
with each ``twisted complex,'' $M_i \in {\rm hmf}_{W_i}$,
we can associate a Landau-Ginzburg model and a D-brane in it.
The tensor product of these Landau-Ginzburg
models is a two-dimensional $\CN=2$ theory, {\it cf.}
\refs{\Emanuel,\Emanuelii,\HoriJW,\Brunneriii}:
\eqn\prodlg{ \otimes_i~ {\rm LG}_{W_i} }

We refer the reader to
\refs{\Khovanov,\Khovanoviii,\RKhovanov,\DBN}
for further details on the construction
of the $sl(N)$ knot homology.


\newsec{Physical Interpretation of the $sl(N)$ Knot Cohomology}

In this section we start by briefly reviewing\foot{For a more detailed
review on these topics the reader can consult \mirbook.}
the relations between  BPS states
and topological strings.  We then discuss how Chern-Simons
theory on $M$ is embedded in open topological strings on $T^*M$.  We  also
review how in the case of $M=\S^3$ `large N' stringy
dualities relate it to closed topological string on a deformed geometry (the
small resolution of the nodal singularity).  Also reviewed is
how the HOMFLY polynomial invariants
can be formulated in this context.  We then go on to define
a refined version of the BPS degeneracies which leads to a refinement
of open topological string amplitudes.  We then propose a relation between
this refined BPS degeneracies
 to the $sl(N)$ knot homology which is a categorification of quantum
$sl(N)$ polynomial invariants.

\subsec{Topological Strings and BPS Degeneracies}

Closed topological string, which encodes Gromov-Witten invariants,
deals with holomorphic maps from closed Riemann surfaces
of arbitrary genus to Calabi-Yau threefolds $X$.
The partition function $Z$ is a function of
the K\"ahler moduli of Calabi-Yau threefold, which we denote
by $T$ and of the string coupling constant $g_s$.  The dependence
on $T$ arises by weighing each map by $\exp(-A)$ where $A$ is the area of the curve.
The $g_s$ dependence enters by weighing a genus $g$ Riemann surface by
the factor $g_s^{2g-2}$.  It has been argued in \GViii\ that $Z$ can
be rewritten in terms of {\it integers} $n^k_{Q}$ as
$$Z(T,g_s)=\prod_{k,Q} \Big[ \prod_{n=1}^{\infty}
 (1-q^{2n+2k} e^{-\langle Q, T\rangle})^n \Big]^{n^k_Q}$$
where $q=e^{-g_s/2}$ and $Q$ denotes an element of $H_2(X,\Z)$.  The integers
$n^k_Q$ denote the degeneracy of BPS states.  More precisely these
correspond to embedded holomorphic curves in the class $Q$ and the index
$k$ is related to the genus of the curve.  Physically these degeneracies
are computed by studying the moduli spaces of D-branes in the class $Q$.
Note that the moduli space includes the choice of a flat connection on the
D-brane in addition to the moduli associated to moving the D-brane inside $X$.
Let us consider an idealized situation
(the more general case was discussed in \KKV ; see also
 \refs{\HSTa,\HSTb,\MNOP,\Katz}):
Suppose that as we vary the curve inside $X$ its genus
does not change (and assume there are no monodromies of the cycles
of the Riemann surface).  In this case the moduli space naturally
splits into the moduli space of the flat connection, which is
a $2g$ dimensional torus $T^{2g}$, and that of the
geometric moduli ${\cal M}^Q_{Geom.}$:
$${\cal M}^Q=T^{2g}\times {\cal M}_{Geom.}$$
In this simple situation one finds that
$$\sum_k n^k_Q q^k=\chi ({\cal M}_{Geom.}) \cdot (q-q^{-1})^{2g}$$
Here the factor $(q-q^{-1})^{2g}$ corresponds to the Poincar\'e polynomial
of $T^{2g}$.  The reader may wonder why the Poincar\'e polynomial
for the full space ${\cal M}^Q$ does not enter the topological string
amplitudes, and only the Euler characteristic of the geometric moduli enters?
The reason is that as one changes the complex moduli of Calabi-Yau $X$, it turns
out that ${\cal M}_{Geom.}$ changes topology (and even dimension!), but
its Euler characteristic does not change.  So only $\chi$ is an invariant
of the Calabi-Yau $X$.

However, there are cases where $X$ is rigid and has no complex moduli.
In such cases one could have obtained a more refined invariant by introducing
a new parameter $t$ which captures the Poincar\'e polynomial $P_{\CM_{Geom}}(t)$
 of the geometric moduli.
In the physical context this corresponds to an addition `charge' that
we can measure and refine the degeneracies to $n^{k,k'}_Q$.  This extra
charge appears as follows: In compactifications on M-theory on Calabi-Yau 3-folds
we end up with a four dimensional space $\IR^4$.  The rotation symmetry group
is thus $SO(4)=SU(2)_L\times SU(2)_R$.  The charge measured by powers of $q$ (together
with $(-1)^F$)
corresponds to $U(1)_L\in SU(2)_L$.  We could also measure the charge $U(1)_R\in SU(2)_R$.
So we would have considered
$$\sum_{k,k'} n^{k,k'}_Q q^{2k} t^{k'}=(q-q^{-1})^{2g} P_{\CM_{Geom}}(t)$$
In fact in view of the application to the knots it is natural
to redefine the basis of the $U(1)_L\times U(1)_R$ charges.  This
is because in that case we have to introduce Lagrangain D-branes which
fill $\IR^2\subset \IR^4$.  We still have a $U(1)^{\otimes 2}$ symmetry.
However one of the $U(1)$'s corresponds to the physical spin
in the subspace $\IR^2$ and is given by the diagonal in $U(1)_L\times U(1)_R$.
We will thus redefine\foot{The choice of $i=\sqrt{-1}$ and also the factor
of $1/2$ in the exponent of $t$ in the formula is for convenience
of comparison with the conventions used in the knot theory literature. }
$$q\rightarrow iqt^{\half} \ $$
and write
\eqn\bpsc{P_Q(q,t)=\sum_{k,k'} n^{k,k'}_Q 
(q^2t)^k (t/q^2)^{k'}=(-1)^g (qt^{\half}+q^{-1} t^{-\half})^{2g} P_{\CM_{Geom}}(t)}
This structure of course presupposes that the moduli space ${\cal M}^Q$ has a product
structure.  This is seldom the case.  One case where this happens naturally is
when the geometric moduli space is a point.  One may think that in the more general
case the splitting of the Poincar\'e polynomial to two variable polynomials
may not exist.
However physical reasoning, i.e. the existence of {\it two} $U(1)$ charges that one
can measure, guarantees that this must be possible {\it even if the  moduli
space does not split}.  Moreover if $X$ is rigid it is an invariant for the Calabi-Yau.
Examples of this have already been computed and checked against physical
computations in \IqbalV, see also \refs{\HSTb,\SchwarzS}
for a mathematical discussion.
This means that $P_Q(q,t)$ always exists but it does not in general take
the simple form given above as a product of two prefactors.

One can also consider open topological strings.  To do this
we consider Lagrangian subspaces $\CL_i$ of $X$ with some multiplicity $N_i$
associated to each.  We then consider holomorphic maps from Riemann
surfaces with arbitrary boundaries such that the boundaries all lie on
some Lagrangian brane $\CL_i$.  Each such configurations will receive
a factor $N_i$; or if the boundary circle wraps over the cycle $\S^1$ $r$ times,
 with holonomy factor $U_i$, we pick up a factor of $tr U_i^r$.  The open topological
 string amptitudes can again be recast in terms of the BPS degeneracies, D2 branes
 embedded in $X$ which can have a boundary on the Lagrangian submanifold.  The degeneracies
will also have a label of Representation of the brane, encoding different ways the
D2 branes end on the Lagrangian submanifold.  The story in this case is somewhat
more complicated that in the closed string case \refs{\OV ,\LMV} .  However
just as in the closed string case one expects the counting of these degeneracies
to depend on choice of charge $Q$ in the relative homology $H_2(X,\CL_i)$, the spin $s$
and in addition on the
representation label $R$, $N_{R,Q,s}$, which is related to cohomology computations on the moduli
space of D2 branes with boundaries (again including the flat connection).  The open
topological string amplitudes can be written in terms of $N_{R,Q,s}$ as discussed in
\fcsviaf .   

What we would like to ask now is whether we can define a more refined invariant even
in this case, in other words, are there more physical charges we can measure?
Let us recall how the open string BPS charges get embedded in the closed
string charges \refs{\OV, \LMV}:  As noted above we have Lagrangian D-branes
filling $\IR^2\subset \IR^4$ and thus in the open string context we {\it still} 
have both $U(1)$ symmetries. Let us denote
these two charges by $(s,r)$, where $s$ denotes the spin in $\IR^2$.   
Thus we expect that any degeneracies
can also be further refined according to these two charges, in particular we would
get new degeneracies $\tilde D_{R,Q,s,r}$ which have the property that
\eqn\ndrefineda{ N_{R,Q,s}=\sum_r (-1)^r \tilde D_{R,Q,s,r} }
One could ask if the integers $\tilde D_{R,Q,s,r}$ are invariant, i.e. do
they depend on moduli of Calabi-Yau or moduli of the brane?  It is easy
to see, just as in the closed string case, that they could depend on the complex
structure moduli of the Calabi-Yau.  But they cannot depend
on the K\"ahler moduli of the Calabi-Yau, or the moduli of the brane.  The latter statement
follows from the fact that the BPS states are in the $(c,c)$ multiplets
whereas the K\"ahler moduli of Calabi-Yau or the brane moduli  are in the $(a,c)$ multiplets
and so it cannot give mass to the $(c,c)$ multiplets.  In particular if we
are dealing with a rigid Calabi-Yau we would find the more refined
invariants $\tilde D_{R,Q,s,r}$, similar to what we discussed for the closed string BPS states.

In this paper for simplicity we will
mainly concentrate on the case where $R=\tableau{1}$, i.e., $N_{\tableau{1},Q,s}$
and the corresponding refinement will be denoted by $\tilde D_{Q,s,r}$.
For the representation $R=\tableau{1}$ the only relevant BPS brane
are Riemann surfaces with one boundary on the Lagrangian submanifold.
Suppose we have an isolated Riemann surface of genus $g$ with one boundary.
Then, just as we discussed in the closed string case we would find
the refined Poincar\'e polynomial:
\eqn\openbps{
\sum_{s,r} \tilde D_{Q,s,r} q^s t^r = (-1)^g (qt^{\half}+q^{-1} t^{-\half})^{2g}
}

\subsec{Chern-Simons Theory and Open Topological Strings}
 
 It was shown in \Wittencsstring\ that $U(N)$ Chern-Simons theory on a three manifold
 $M$ can be reformulated in terms of {\it open} topological string on the Calabi-Yau
 manifold $X=T^*M$, where we consider $N$ D-branes wrapping the Lagrangian
 cycle $M$.  What this means is that we consider Riemann surfaces with boundaries
 and `count' holomorphic maps from the Riemann surface onto $T^*M$ with the condition
 that the boundary ends on $M$, and that each boundary gets a factor of $N$ corresponding
 to which brane it ends on.  Moreover the Chern-Simons coupling $2\pi i/(k+N)$ gets
 identified with topological string coupling constant $g_s$.  In fact there are no
 honest holomorphic maps from Riemann surfaces to $T^*M$ however we end up with
 degenerate maps which approach a holomorphic map.  These degenerate maps
 correspond to ribbon graphs on $M$ and in fact reproduce the Feynman graphs
 of the Chern-Simons perturbation
 theory.
 
\subsec{Open Topological String on $\S^3$ and a Large N Duality}
 
It was shown in \GopakumarV\ that for the case of $M=\S^3$, with $N$ branes
wrapping $\S^3$ there
is an equivalent description of the {\it open} topological strings in terms of
a {\it closed} topological string.  The corresponding Calabi-Yau is obtained
by a geometric transition:  $\S^3$ shrinks to a point, leading to conifold
which can be smoothed out by a small blow up leading to a Calabi-Yau
geometry which is a total space of the bundle $\CO (-1)\oplus \CO (-1)$ over
${\bf P}^1$.  Moreover the K\"ahler moduli of ${\bf P}^1$
(i.e. its area) is $T=N g_s=2\pi iN/k+N$. In other words
$${\exp }(-T/2)=\lambda =q^N$$
This therefore gives a relation between Chern-Simons theory on $\S^3$ and closed
topological strings on the resolved conifold.

\subsec{Introducing the Knots}

One can extend the above relation between Chern-Simons theory
on $\S^3$ and topological strings on resolved conifold, to include
knot observables \OV :  Consider a knot $K$ in $\S^3$.  $K$ defines
a non-compact Lagrangian submanifold $\CL_K$ in $T^* \S^3$ given by the conormal
bundle.  If we wrap $M$
D-branes on $\CL_K$ we also get a Chern-Simons
theory on it.  We have two Wilson loop observables $U,V$ on $\S^3 $ and
$\CL_K$.  Viewing $V$ as fixed, wrapping branes on $\CL_K$ deforms the Chern-Simons
theory on $\S^3$ and gives a new partition function: 
$$ Z(V,N,k)=\sum_R  {\rm tr}_R V \langle {\rm tr}_R U\rangle
=\sum_R ({\rm tr}_R V) W_R(K)$$

The large $N$ duality relates this to computation of topological strings
on the resolved conifold where the branes wrapping $\S^3$ have disappeared, but
the branes wrapping $\CL_K$ continue to be there.  The fact that there should
be such a Lagrangian $\CL_K$ even after the transition has been shown 
in \Taubes .  This means that we can rewrite $Z(V,N,k)$ in terms
of open topological strings in this new geometry.  The $N$ and $k$ dependence
capture the K\"ahler moduli dependence and the string coupling constant, and
the $V$ dependence captures how the Riemann surface ends on $\CL_K$ (if a
boundary wraps around the $\S^1\subset \CL_K$ $r$ times it leads to a factor
$tr V^r$).  In this case the Calabi-Yau is rigid and so we can define
the refined invariants $\tilde D_{R,Q,s,r}$ associated to each knot $K$.
For the case of $R=\tableau{1}$ it is natural to expect this is related to $sl(N)$
knot homology.  Let us try to see if we can make this connection more precise.

{}From the definition of open topological string amplitude and its relation
to quantum $sl(N)$ knot invariant \pnvian\ it follows that
\eqn\pnvianx{P_N (q)= {1 \over q - q^{-1}} \sum_{Q,s,r \in \Z}
(-1)^r q^{NQ+s} \tilde D_{Q,s,r} }
where we also used \ndrefineda.
It is tempting in the above formula to replace
$(-1)$ by a new variable $t$ and write the generating
function of the refined BPS degeneracies $\tilde D_{Q,s,r}$
in the form
\eqn\khviansax{
{1 \over q-q^{-1}} \sum_{Q,s,r \in \Z} \tilde D_{Q,s,r} q^{NQ+ s} t^r
}
We expect, that there is a similar expression for
the graded Poincar\'e polynomial of the $sl(N)$ homological invariant
in terms of some integers $D_{Q,s,r}$, {\it cf.} \khvians:
\eqn\khviansa{
Kh_N (q,t) = {1 \over q-q^{-1}} \sum_{Q,s,r \in \Z} D_{Q,s,r} q^{NQ+ s} t^r
}
where $D_{Q,s,r}$ obey
\eqn\ndrefined{
N_{\tableau{1},Q,s} = \sum_{r \in \Z} (-1)^r D_{Q,s,r}
}
The above statement, if true, would give a highly non-trivial
prediction for the $N$-dependence of the homological $sl(N)$ invariants.
In fact, we find this dependence is satisfied in all the examples
we have checked, as will be discussed in the next section.

Although we do not claim that the generating functions \khviansax\
and \khviansa\ need to be equal, we expect a simple relation between
the refined BPS invariants, $D_{Q,s,r}$, and their physical
counterparts, $\tilde D_{Q,s,r}$. Indeed, in the examples
considered below, one can recognize contributions
of genus-$g$ curves of the form
\eqn\tqcontrib{
\sum_{s,r} D_{Q,s,r} q^s t^r
= (-1)^g (qt^{\half}+q^{-1} t^{-\half})^{2g} t^{\alpha_K Q}
}
where $\alpha_K$ is some simple invariant of the knot $K$.
Comparing this expression with \bpsc\ and \openbps,
one can interpret such terms either as contributions
of isolated curves, in which case $P_{\CM_{Geom}} (t)= 1$,
or as contributions\foot{Such contributions
can arise if there is a non-trivial monodromy.}
of genus-$g$ curves with $P_{\CM_{Geom}} (t) = t^{\alpha_K Q}$.
In the first case, we expect the extra factor of $t^{\alpha_K Q}$
to come from the change of basis for $(Q,s,r)$ in the relation
between $D_{Q,s,r}$ and $\tilde D_{Q,s,r}$:
\eqn\ddrelone{
D_{Q,s,r} = \tilde D_{Q,s,r - \alpha_K Q } }
This change of basis would be very natural in a relation
between two graded homology theories. Notice, it does not
affect the $N$-dependence of \khviansax\ - \khviansa\ which is
true even without such a shift. It would be important to understand
this change of basis more deeply. It is tempting to speculate
that this is related to replacing the $sl(N)$ knot homology with
a $gl(N)$ version, as the corresponding Chern-Simons theory
related to topological strings is the $U(N)$ version and not the $SU(N)$
version.

On the other hand, if we interpret \tqcontrib\ as a contribution
of genus-$g$ curve with $P_{\CM_{Geom}} (t) = t^{\alpha_K Q}$,
there is no need for the change of basis $r\rightarrow r-\alpha_K Q$,
and we can simply identify $D_{Q,s,r} = \tilde D_{Q,s,r}$.



%


\newsec{Examples}

\subsec{The Unknot}

The conjecture \khvians\ automatically holds for the unknot, the only
knot for which all the $sl(N)$ knot homology groups are known at
present.

Indeed, for the unknot, the $sl(N)$ knot cohomology
coincides with its graded Euler characteristic \njonesunknot\
(since all the non-trivial groups $\CH_N^{i,j} ({\rm unknot})$ are
only in degree $i=0$). Therefore, the graded Poincar\'e polynomial
manifestly obeys the relation \khvians,
$$
Kh_N (q,t) = {1 \over q-q^{-1}}
\sum_{Q,s,r} D_{Q,s,r} q^{NQ + s} t^r
$$
with non-zero invariants
\eqn\dforunknot{ D_{-1,0,0} = -1, \quad D_{1,0,0} = 1 }
Notice, that, in the case of the unknot, the moduli
spaces $\CM_{g,Q}$ are isolated points when $Q=1$ and $Q=-1$,
{\it cf.} \hgvunknot. These are precisely the values of $Q$
for which we find non-trivial invariants \dforunknot.
Moreover, for each value of $Q$, the invariants \dforunknot\
have the structure \openbps\ consistent with a contribution
of an isolated curve with $g=0$.

Let us consider the explicit form of the $sl(N)$ cohomology
groups for the unknot. For example, for $N=4$ we have
\eqn\khunknot{
\CH_{N=4}^{i,j}: \quad \matrix{
0 & 0 & 0 & 0 & 0 & 0 & 0 & 0 & 0 & 0 & 0 & 0  & 0 \cr
0 & 0 & 0 & 0 & 0 & 0 & 0 & 0 & 0 & 0 & 0 & 0  & 0 \cr
0 & 0 & 0 & \IQ & 0 & \IQ & 0 & \IQ & 0 & \IQ & 0 & 0 & 0
}}
where the vertical direction represents the index $i$,
while the horizontal direction represents the grading $j$.
In general, the complex $\CH_N^{\bullet} ({\rm unknot})$
has $N$ non-trivial elements.
On the other hand, the complex \khshifted\ obtained from \khunknot\
has only two non-trivial elements,
\eqn\unknotxxb{\quad\quad\quad\quad
\matrix{
0 & 0 & 0 & 0 & 0 & 0 & 0 & 0 & 0 & 0 & 0 & 0 & 0 \cr
0 & 0 & 0 & 0 & 0 & 0 & 0 & 0 & 0 & 0 & 0 & 0 & 0 \cr
0 & 0 & \ominus \IQ & 0 & 0 & 0 & 0 & 0 & 0 & 0 & \IQ & 0 & 0
}}
where we used the equivalence $\IQ \ominus \IQ \cong 0$.


\subsec{The Trefoil Knot}

In this case, the complex \khshifted\ has six non-trivial
elements\foot{For example, for $N=2$ it has the form
\eqn\xxtrefoil{
\matrix{
3 ~\vert & & & & & & & & & \ominus \IQ & & \IQ &  \cr
2 ~\vert & & & & & \ominus \IQ & & \IQ & & & & &  \cr
1 ~\vert & & & & & & & & & & & &  \cr
0 ~\vert & \ominus \IQ & & & & \IQ & & & & & & &  \cr
~~+ & -0- & -1- & -2- & -3- & -4- & -5- & -6- & -7- & -8- & -9- & -10- &  \to j
}}
where the vertical (resp. horizontal) direction represents
index $i$ (resp. index $j$) and we used $\IQ \ominus \IQ \cong 0$.}.
Similarly, there are only six non-zero BPS
invariants $N_{\tableau{1},Q,s}$ for the following values of $(Q,s)$:
\eqn\qstrefoil{ (1,-2), \quad (3,-2), \quad (1,2), \quad (3,0),
\quad (3,2), \quad (5,0) }
It is natural to expect that each of the non-zero BPS invariants
comes from a single cohomology group which, in turn,
corresponds to a particular element in \khshifted\ with the same
value of $(Q,s)$:
\eqn\ndrefinednosum{
N_{\tableau{1},Q,s} = (-1)^r D_{Q,s,r}
}
Notice, there is no summation over $r$ in this formula.
There is a unique choice of such $D_{Q,s,r}$
consistent with the $sl(2)$ homological invariant \khtrefoil:
%
\eqn\guesstrefoil{\eqalign{
& D_{3,-2,0} =1, \quad D_{3,0,2} = 1, \quad D_{5,0,3} = 1, \quad \cr
& D_{1,-2,0} = -1, \quad D_{1,2,2} = -1, \quad D_{3,2,3} = -1 \quad  }}
Given these numbers, one can use \khvians\ to compute all $sl(N)$
cohomology groups for the trefoil knot. For example, substituting
\guesstrefoil\ into \khvians\ we find the following homological
$sl(3)$ invariant:
\eqn\slthreetrefoil{ Kh_{N=3} (3_1) = q^2 + q^4 + q^6 + q^6 t^2 +
q^8 t^2 + q^{12} t^3 + q^{14} t^3 }
Remarkably, this result is in perfect agreement with the one
obtained using the technique of foams\foot{We thank
M.~Khovanov for explanations and very helpful discussions on
the ``computational shortcuts'' used in these computations.} \Khovanoviii.
For the $sl(4)$ and $sl(5)$ homology, we predict
\eqn\slfourtrefoil{ Kh_{N=4} (3_1) = q^3 + q^5 + q^7 + q^9 + q^7
t^2 + q^9 t^2 + q^{11} t^2 + q^{15} t^3 + q^{17} t^3 + q^{19} t^3
}
\eqn\slfivetrefoil{ Kh_{N=5} (3_1) = q^4 + q^6 + q^8 + q^{10} +
q^{12} + q^8 t^2 + q^{10} t^2 + q^{12} t^2 + q^{14} t^2 + q^{18}
t^3 + q^{20} t^3 + q^{22} t^3 + q^{24} t^3 }

We note that, the structure of the refined integer
invariants \guesstrefoil\ for $Q=1$ and $Q=5$ has
the form \tqcontrib\ consistent with a contribution of
genus-$g$ curves. For example, for $Q=1$ we have
\eqn\gcontrtrefoil{ \sum_{s,r} D_{Q,s,r} q^s t^r
= t \Big[ -(q^{-1} t^{-1/2} + q t^{1/2})^2 +2 \Big] }
which corresponds to a contribution of curves
with $g=0$ and $g=1$. This is a strong evidence
suggesting that embedding the knot homology in
our physical setup is indeed correct.
Moreover, since both terms in \gcontrtrefoil\ have an extra factor of $t$,
it also suggests, though not conclusively, that these curves
are isolated and that  we indeed have an affine change of basis
in comparing knot homology with BPS spectrum, as discussed in the previous section.
It would be extremely interesting to consider
the Lagrangian D-brane associated to the trefoil
and check whether or not there are isolated
curves with $Q=1$ and $g=0,1$ ending on it.


We find similar contributions of genus-$g$ curves for other
torus knots, $T_{2,2m+1}$, where the maximal genus of such
curves grows linearly with the number of crossings, $g_{max} = m$.
Again, these examples suggest that, in order to compare the invariants
$D_{Q,s,r}$ with the physical BPS degeneracies $\tilde D_{Q,s,r}$,
one has to make a change of basis, $r\rightarrow r-{Q/2}$.
Note that in the case of the unknot we did not have to change the basis.
This suggests that the change of basis involves a simple knot invariant
$\alpha_K$, {\it cf.} \ddrelone.


\subsec{The Knot $5_1$}

Following the same reasoning as in the
previous example, we consider the set of values $(Q,s)$ which
correspond to non-zero integer invariants $N_{\tableau{1},Q,s}$:
\eqn\qsfiveone{ (3,0), \quad (3,-4), \quad (3,4), \quad (5,0),
\quad (5,-2), \quad (5,2), \quad (5,-4), \quad (5,4), \quad
(7,-2), \quad (7,2) }
Again, following the natural assumption that these are the values
of $(Q,s)$ which correspond to non-trivial elements in \khshifted,
we can find the invariants $D_{Q,s,r}$ which encode
the $sl(N)$ homological invariants,
\eqn\guessfiveone{\eqalign{
& D_{5,-4,0} = 1, \quad D_{5,-2,2} = 1, \quad D_{7,-2,3} = 1, \quad
D_{5,2,4} = 1, \quad D_{7,2,5} = 1, \quad \cr
& D_{3,-4,0} = -1, \quad D_{3,0,2} = -1, \quad D_{5,0,3} = -1, \quad
D_{3,4,4} = -1, \quad D_{5,4,5} = -1 }}
Given these numbers, one can use \khvians\ to compute all $sl(N)$
cohomology groups for the knot $5_1$. For example, in the case
$N=3$ we get
\eqn\slthreefiveone{
Kh_{N=3} (5_1) = q^6 + q^8 + q^{10} + q^{10} t^2 + q^{12} t^2 +
q^{16} t^3 + q^{18} t^3 + q^{14} t^4 + q^{16} t^4 + q^{20} t^5 + q^{22} t^5 }
Again, this result is in a perfect agreement with the $sl(3)$ knot
homology computed using the technique of foams \Khovanoviii.
It would be interesting to check the other $sl(N)$ knot homology
groups that follow from \guessfiveone.

It is easy to generalize these examples to arbitrary
torus knots $T_{2,2m+1}$. For every torus knot $T_{2,2m+1}$,
we find a complete agreement between our predictions based on \khvians\
and the $sl(3)$ knot homology groups computed using the technique
of web cobordisms (the details will be discussed elsewhere \DGR).


\newsec{Generalizations}

The proposed relation between the knot cohomology
and the spectrum of BPS states leads to non-trivial predictions
for both knot theory and physics.
Apart from the computational predictions, examples of which we
discussed in the previous section,
there are interesting generalizations suggested by this relation.
For example, it suggests that there should exist a categorification
of more general knot invariants associated with arbitrary
representations of $sl(N)$, not just the fundamental representation.
Similarly, since knot cohomology can be defined over
arbitrary ground fields, including finite number fields,
there should exist corresponding physical realizations
in terms of BPS states.
This would be interesting to study further.
Also, it would be very exciting to find a combinatorial
definition of the integer invariants $D_{Q,s,r}$,
and of their generalizations to other representations.

The formulation of the $sl(N)$ homological invariant
in terms of the refined BPS invariants $D_{Q,s,r}$
can also be extended to links.
Let $L$ be an oriented link in $\S^3$
with $\ell$ components, $K_1, \ldots, K_{\ell}$,
all of which carry a fundamental representation of $sl(N)$.
As in \pnjones, the expectation value of the corresponding Wilson loop
operator $W (L) = W_{\tableau{1}, \cdots, \tableau{1}} (L)$
is related to the polynomial $sl(N)$ invariant,
with a minor modification,
\eqn\pnlinks{ P_N (L) = q^{-2N \lk (L)} \langle W (L) \rangle }
where $\lk (L) = \sum_{a<b} \lk (K_a,K_b)$
is the total linking number of $L$.
The formulation of $P_N (L)$ in terms of integer
BPS invariants also needs some modification \LMV.
Namely, in our notations,
\eqn\wcvian{
\langle W (L) \rangle^{(c)} = (q^{-1}-q)^{\ell-2}
\sum_{Q,s} N_{(\tableau{1}, \cdots, \tableau{1}),Q,s} q^{NQ+s}
}
where $\langle W (L) \rangle^{(c)}$
is the {\it connected} correlation function.
For example, for a two-component link, we have
\eqn\wconnected{
\langle W (L) \rangle^{(c)} =
\langle W (L) \rangle - \langle W (K_1) \rangle \langle W (K_2) \rangle
}
and
\eqn\pntwocomp{
P_N (L) = q^{-2N \lk (L)} \Big[ P_N (K_1) P_N (K_2)
+ \sum_{Q,s} N_{(\tableau{1}, \tableau{1}),Q,s} q^{NQ+s}
\Big] }
where $P_N (K_1)$ and $P_N (K_2)$ denote the $sl(N)$ polynomials
of the link components.

We wish to write a similar expression for the graded Poincar\'e
polynomial $Kh_N (L)$ in terms of integer invariants $D_{Q,s,r} (L)$.
Because of the corrections involving the $sl(N)$ invariants
of the sublinks, this formulation is less obvious
than in the case of knots.
For example, for a two-component link $L$,
we find, {\it cf.} \khvians,
\eqn\khtwocompviad{
Kh_N (L) = q^{-2N \lk (L)} \Big[
t^{\a} Kh_N (K_1) Kh_N (K_2) + {1 \over q-q^{-1}}
\sum_{Q,s,r \in \Z} D_{Q,s,r} q^{NQ+s} t^r \Big]
}
where $\a$ is a simple invariant of $L$,
and the integer invariants $D_{Q,s,r}$
and $N_{(\tableau{1}, \tableau{1}),Q,s}$
are related as follows:
\eqn\nslviad{ N_{(\tableau{1}, \tableau{1}),Q,s-1}
- N_{(\tableau{1}, \tableau{1}),Q,s+1}
= \sum_{r \in \Z} (-1)^r D_{Q,s,r}
}
Notice, this relation is quite different from what
we had in the case of knots, {\it cf.} \nqsviad.

The first term in \khtwocompviad\ is similar to
the first term in \pntwocomp.
In order to understand the structure of
the second term, note that
the factor $(q^{-1}-q)^{\ell-2}$ in \wcvian\ is a product of
two terms, $(q^{-1}-q)^{-1}$ and $(q^{-1}-q)^{\ell-1}$,
which have different origin and should be treated differently.
The first term, $(q^{-1}-q)^{-1}$, comes from the Schwinger
computation \OV\ and remains intact once $P_N (L)$ is lifted
to the homological $sl(N)$ invariant. On the other hand,
the factor $(q^{-1}-q)^{\ell-1}$ is similar to
the contribution of a genus-$g$ curve, $(q^{-1}-q)^{2g}$,
discussed in section 3.
In the homological $sl(N)$ invariant this factor is
replaced by a polynomial expression in $q^{\pm 1}$ and $t^{\pm 1}$.
Applying the same logic to \wcvian\
gives the second term in \khtwocompviad.


\vskip 25pt

\centerline{\bf Acknowledgments}

We would like to thank D.~Bar-Natan, R.~Dijkgraaf, M.~Gross, K.~Intriligator,
A.~Kapustin, M.~Khovanov, A.~Klemm, M.~Mari\~no, H.~Ooguri, J.~Roberts and
D.~Thurston for useful discussions.
S.G. would also like to thank the Caltech Particle
Theory Group for kind hospitality.
The work of A.S. is supported by NSF grant DMS-0204927.
This work was conducted during the period S.G. served
as a Clay Mathematics Institute Long-Term Prize Fellow.
S.G. is also supported in part by RFBR grant 04-02-16880.
The work of C.V. is supported in part by
NSF grants PHY-0244821 and DMS-0244464.

\medskip

\noindent
{\bf Note added:}
A preliminary version of the present work was presented by one of us (S.G.)
at Caltech and UCSD seminars in October 2004.
Since then, there have been
some interesting developments which provide additional support
for the conjecture \khvians, see {\it e.g.} \DBNnews\ for
independent computations of the Khovanov-Rozansky homology
in some of the examples considered here (see also \DGR).

\vfill
\eject

\appendix{A}{HOMFLY Polynomial for Some Knots}

\noindent
Let us list the polynomial invariant $f_{\tableau{1}} (q,\lambda)$
for some simple knots:
$$
\eqalign{
& f_{\tableau{1}} (3_1) = {\lambda q^{-2} + q^2 \lambda
- \lambda^3 - \lambda^3 q^{-2} - q^2 \lambda^3 + \lambda^5 \over q^{-1}-q } \cr
& f_{\tableau{1}} (4_1) = {\lambda^{-3} - \lambda^{-1} q^{-2}
- q^{2} \lambda^{-1} + \lambda q^{-2} + \lambda q^2
- \lambda^{3} \over q^{-1}-q } \cr
& f_{\tableau{1}} (5_1) = { \lambda^3 + \lambda^3 q^{-4} + q^4 \lambda^3
- \lambda^5 - \lambda^5 q^{-4} - \lambda^5 q^{-2} - q^2 \lambda^5
- q^4 \lambda^5 + \lambda^7 q^{-2} + q^2 \lambda^7 \over q^{-1}-q } \cr
& f_{\tableau{1}} (6_1) = { \lambda^{-3} + \lambda^{-1}
- q^{-2} \lambda^{-1} - q^2 \lambda^{-1} - \lambda + \lambda^3 q^{-2}
+ q^2 \lambda^3 - \lambda^5 \over q^{-1}-q }
}
$$
According to \fnhat, the polynomial $f_{\tableau{1}} (q,\lambda)$
can be viewed as a generating function of the BPS invariants
$\hat N_{\tableau{1},g,Q}$ (or $N_{\tableau{1},Q,s}$).
Notice, that the BPS invariants for
the knots $3_1$, $4_1$, and $6_1$ obey a simple relation,
\eqn\tfsrelgv{
\hat N_{\tableau{1},g,Q} (3_1) - \hat N_{\tableau{1},g,Q} (4_1)
+ \hat N_{\tableau{1},g,Q} (6_1) = \hat N_{\tableau{1},g,Q} ({\rm unknot})
}

\appendix{B}{$sl(2)$ Knot Homology for Some Knots}

\noindent
Here, following \DBN, we list Khovanov's $sl(2)$ invariants, $Kh_2 (L)$,
for some simple knots:
$$
\eqalign{
& Kh_2 (3_1) = q + q^{3} + q^{5} t^{2} + q^{9} t^{3} \cr
& Kh_2 (4_1) = q + q^{-1} + q^{-1}t^{-1} + qt + q^{-5} t^{-2} + q^{5} t^{2} \cr
& Kh_2 (5_1) = q^{3} + q^{5} + q^{7} t^{2} + q^{11} t^{3}
+ q^{11} t^{4} + q^{15} t^{5} \cr
& Kh_2 (5_2) = q + q^{3} + q^{3} t + q^{5} t^{2} + q^{7} t^{2}
+ q^{9} t^{3} + q^{9} t^{4} + q^{13} t^{5} \cr
& Kh_2 (6_1) = 2q^{-1} + q^{-1} t^{-1} + q^{-5} t^{-2} + q + q t + q^{3} t
+ q^{5} t^{2} + q^{5} t^{3} + q^{9} t^{4} \cr
& Kh_2 (7_1) = q^{5} + q^{7} + q^{9} t^{2} + q^{13} t^{3}
+ q^{13} t^{4} + q^{17} t^{5} + q^{17} t^{6} + q^{21} t^{7}
}
$$
Notice, that the $sl(2)$ homological invariants
for the knots $3_1$, $4_1$ and $6_1$ are closely related:
\eqn\tfsrelkh{\eqalign{
Kh_2 (3_1) - Kh_2 (4_1) + Kh_2 (6_1)
= & q + q^{-1} \cr
& + q^{3} (1+t) + q^{5} (t^{2}+t^{3})
 + q^{9} (t^{3}+t^{4})
}}
The right-hand side of this expression evaluated at $t=-1$
gives $Kh_2 ({\rm unknot})$, in agreement with \tfsrelgv.

\listrefs
\end